\documentclass[twocolumn,pra,showpacs]{revtex4}
\usepackage{amsmath,amsfonts}
\begin{document}
\draft
\title{\bf Enhanced nuclear Schiff moment and time reversal violation in $^{229}$Th - containing  molecules }
\author{V.V. Flambaum
$^{1,2}$
} 
\affiliation{
$^1$
School of Physics, University of New South Wales,  Sydney 2052,  Australia}
\affiliation{$^2$Johannes Gutenberg-Universit\"at Mainz, 55099 Mainz, Germany}
\date{\today}

\begin{abstract}
Octupole deformation results in a strongly enhanced collective Schiff moment  in $^{229}$Th nucleus. An additional enhancement of time reversal (T) and parity (P) violating effects (such as T,P-violating electric dipole moments) appears  in the ground $^1\Sigma$ state and in the metastable  $^3\Delta_1$ state of diatomic molecule $^{229}$ThO. Similar enhancements exist in molecular ions  $^{229}$ThOH$^+$,  $^{229}$ThF$^+$ and $^{225,223}$RaOH$^+$. Corresponding experiments may be used to test CP-violation theories predicting T,P-violating nuclear forces and to search for axions.
 \end{abstract}

\maketitle
{\bf Introduction: octupole deformation and enhanced nuclear Schiff moments}.  Measurements  of  T,P and CP -violating electric dipole moments (EDM) of elementary particles, nuclei and atoms  provide crucial tests of unification theories and have already cornered many popular models of CP-violation including supersymmetry \cite{PR,ERK}. Corresponding effects are very small, therefore, we are looking for the enhancement mechanisms - see e.g.  \cite{Khriplovich,KL,GF}. 

According to the Schiff theorem the nuclear EDM is completely screened  in neutral atoms  \cite{Schiff}.  EDM of diamagnetic atoms is produced by interaction of electrons with the nuclear Schiff moment.  The Schiff moment is a vector multipole which produces electric field nside the nucleus. It  appears in the third order of the multipole expansion of the  nuclear electrostatic potential with added electron screening term 
 \cite{Sandars,Hinds,SFK,FKS1985,FKS1986}.  The distribution of the Schiff moment electric field inside the nucleus, the Hamiltonian describing interaction of the Schiff moment with relativistic atomic electrons  and the  finite nuclear size corrections to the formula for the Schiff moment have been considered in Ref. \cite{FG}.

Refs. \cite{Sandars,Hinds} calculated  the Schiff moment due to proton EDM.   Refs. \cite{SFK,FKS1985,FKS1986} calculated (and named) the nuclear Schiff moment  produced by the P,T-odd nuclear forces. It was shown in \cite{SFK} that the contribution of the P,T-odd forces to the nuclear EDM and Schiff moment is $\sim 40$ times larger than the contribution of the nucleon EDM.   

    Further enhancement of the nuclear Schiff moment may be due to the close nuclear levels of opposite parity  with the same angular momentum which can be mixed by the T,P-odd nuclear forces \cite{SFK} (nuclear EDM and magnetic quadrupole can also be  enhanced \cite{HH}).  Nuclear T,P-odd moments such as magnetic quadrupoles may also be enhanced due to their collective nature in deformed nuclei \cite{F1994}. However, the largest enhancement ($\sim 10^2 - 10^3$ times) happens in nuclei with the octupole deformation where both the close nuclear level effect and the collective effect work together \cite{Auerbach,Spevak}.
    
{\bf Calculation of collective Schiff moment}.  The Schiff moment is defined by the following expression \cite{SFK}: 
\begin{equation}\label{S}
{\bf S}=\frac{e}{10} [<r^2 {\bf r}> - \frac{5}{3Z}<r^2><{\bf r}>], 
\end{equation}
where $<r^n> \equiv \int \rho({\bf r}) r^n d^3r$ are the moments of the nuclear charge density $\rho$. The second term originates from the electron screening and contains  nuclear mean squared charge radius  $<r^2>/Z$ and nuclear EDM  $d=e<{\bf r}>$, where $Z$ is the nuclear charge. 

If a nucleus has an octupole deformation $\beta_3$ and a quadrupole deformation $\beta_2$, in the fixed-body  (rotating) frame the Schiff moment $S_{intr}$ is proportional to the octupole moment $O_{intr}$, i.e. it  has a collective nature  \cite{Auerbach,Spevak}:
\begin{equation}\label{Sintr}
 S_{intr} \approx \frac{3}{5 \sqrt{35}} O_{intr} \beta_2 \approx    \frac{3}{20 \pi  \sqrt{35}} e Z R^3 \beta_2 \beta_3 ,
\end{equation}
where $R$ is the nuclear radius.
 However, in the laboratory frame EDM and  Schiff moment are forbidden by the parity and   time reversal invariance. Indeed, EDM and Schiff moment are polar $T$-even vectors which must be directed along the nuclear spin $I$ which is $T$-odd pseudovector.  
   
   A  nucleus with an octupole deformation and non-zero nucleon angular momentum  has a doublet of close opposite parity rotational states $|I^{\pm}>$ with the same angular momentum $I$ ($| I^{\pm} >=\frac{1}{\sqrt{2}} (|\Omega> \pm |-\Omega>)$, where $\Omega$ is  the projection of $I$ on to the nuclear axis). The states of of this doublet are mixed by $P,T-$violating interaction $W$. The mixing coefficient is:
\begin{equation}\label{alpha}
 \alpha=\frac{<I^-| W| I^+>}{E_+  -  E_-} . 
\end{equation}
This mixing polarises  nuclear axis ${\bf n}$ along the nuclear spin ${\bf I}$,  $<n_z>= 2 \alpha \frac{I_z}{I+1}$,
and the intrinsic Schiff moment shows up in the laboratory frame \cite{Auerbach,Spevak}:
\begin{equation}\label{Scol}
 S= 2 \alpha \frac{I}{I+1} S_{intr}. 
\end{equation}
A nucleus with an octupole deformation also has a small intrinsic EDM $D$ due to a difference between the proton and neutron distributions which results in the laboratory frame nuclear EDM  $d=2 \alpha \frac{I}{I+1} D$  \cite{Auerbach,Spevak}.

A similar $\Omega$- doublet mixing mechanism produces huge enhancement of electron  EDM $d_e$  and  T,P-odd  interactions in polar molecules, such as ThO. Interaction of $d_e$ with molecular electric field produces the mixing coefficient $\alpha$ resulting in the orientation of  large intrinsic molecular EDM $D \sim e a_B$ along the molecular angular momentum ${\bf J}$, and we obtain  $d=2 \alpha \frac{J}{J+1} D \sim \alpha e a_B$  \cite{SushkovFlambaum}, where $a_B$ is the Bohr radius. As a result,  the $T,P$- violating molecular EDM $d$ exceeds electron EDM $d_e$ by 10 orders of magnitude.   

   In the papers \cite{Auerbach,Spevak}  the numerical calculations of the Schiff moments and estimates of atomic EDM produced by electrostatic interaction between electrons and these moments have been done  for  $^{223}$Ra, $^{225}$Ra, $^{223}$Rn, $^{221}$Fr, $^{223}$Fr, $^{225}$Ac and $^{229}$Pa. The Schiff moment of  $^{225}$Ra exceeds the Schiff moment of $^{199}$Hg (where the most accurate  measurements of the Schiff moment have been performed \cite{HgEDM}) 200 times. Even larger enhancement of the  $^{225}$Ra Schiff moment has been obtained  in Ref. \cite{EngelRa}. For other nuclei the enhancement factors relative to Hg are between 30 and 700.   
  Atomic calculations of  EDM induced by the Schiff moment in Hg, Xe, Rn, Ra and Pu atoms have been performed in Refs. \cite{SFK,FlambaumRa,DzubaRa,AtomicSchiff,AtomicSchiff2} and include additional atomic enhancement mechanisms. 
  
   It is useful to make an analytical estimate of the Schiff moment.  According to Ref. \cite{Spevak} the T,P-violating matrix element is approximately equal to
   \begin{equation}\label{W}
   <I^-| W| I^+> \approx \frac{\beta_3 \eta}{A^{1/3}}  \textrm{eV}.
  \end{equation}  
  Here $\eta$ is the dimensionless strength constant of the nuclear $T,P$- violating potential $W$:
   \begin{equation}\label{eta}
 W= \frac{G}{\sqrt{2}} \frac{\eta}{2m} ({\bf \sigma \nabla}) \rho ,
   \end{equation}
where $G$ is the Fermi constant, $m$ is the  nucleon mass and $\rho$ is the nuclear number density. Eqs. (\ref{Sintr},\ref{alpha},\ref{Scol},\ref{W}) give the analytical estimate for the  Schiff moment: 
\begin{equation}\label{San}
 S \approx 1. \cdot 10^{-4} \frac{I}{I+1} \beta_2 \beta_3^2 Z A^{2/3} \frac{\textrm{KeV}}{E_-  -  E_+} e \,\eta \, \textrm{fm}^3,
   \end{equation}
This estimate gives $S=280 \, e \,\eta \, \textrm{fm}^3$ for $^{225}$Ra, very close to the result  of the numerical calculation in Ref.  \cite{Spevak} $S=300 \, e \,\eta \, \textrm{fm}^3$. 

The values of the Schiff moments for the nuclei  with octupole deformation listed above vary from 45 to 1000 $10^{-8} e \eta$ fm$^3$   \cite{Spevak}. For spherical nuclei  $^{199}$Hg, $^{129}$Xe, $^{203}$Tl and  $^{205}$Tl, where the Schiff moment measurements have been performed,  the calculations \cite{SFK,FKS1985,FKS1986} give the Schiff moment $S \sim 1 \times 10^{-8} $ $e \eta$ fm$^3$. 
    
  The Schiff moment in Eq. (\ref{San}) is proportional to the squared octupole deformation parameter $\beta_3^2$ which is about $(0.1)^2$. According to Ref. \cite{Engel2000}, in nuclei with a soft octupole vibration mode the squared dynamical octupole deformation  $<\beta_3^2 > \sim (0.1)^2$, i.e. it is the same as the static octupole deformation. This means that a similar enhancement of the Schiff moment may be due to the dynamical octupole effect \cite{Engel2000,FZ,soft2} in nuclei where $<\beta_3>=0$.
  
       Unfortunately, the nuclei with the octupole deformation and non-zero spin have a short lifetime. Several experimental groups have considered experiments with  $^{225}$Ra and $^{223}$Rn. The only published EDM measurement  \cite{RaEDM}  has been done for $^{225}$Ra which has 15 days half-life. In spite of the Schiff moment enhancement the $^{225}$Ra  EDM measurement has not reached yet the sensitivity to the T,P-odd interaction  Eq. (\ref{eta}) comparable to the Hg EDM experiment \cite{HgEDM}. The experiments continue, however,  the instability of $^{225}$Ra and a relatively small number of atoms available  may be a problem.
       
       To have a breakthrough in the sensitivity we need a more stable nucleus and a larger number of atoms.  An excellent candidate is $^{229}$Th nucleus which lives 7917 years and is very well studied in numerous experiments and calculations (this nucleus is the only candidate for the nuclear clock which is expected to have a precision significantly  better than atomic clocks \cite{Peik,Thclock}, has strongly enhanced effects of "new physics" \cite{FlambaumTh,FlambaumTh1} and may be used for a nuclear laser \cite{Tkalya}). $^{229}$Th is produced in macroscopic quantities by the decay of $^{233}$U (see e.g. \cite{production}), and its principal use is for the production of the medical isotopes $^{225}$Ac and $^{213}$Bi.
       
       According to Ref. \cite{Minkov} the $^{229}$Th nucleus has the octupole deformation with the parameters $\beta_3$=0.115,   $\beta_2$=0.240, $I=5/2$ and the  interval between the opposite parity levels $E(5/2^-)- E(5/2^+)$=133.3 KeV.  The analytical formula in Eq. (\ref{San}) allows us to scale the value of the Schiff moment from the numerical calculations  for $^{225}$Ra which has    $\beta_3$=0.099,   $\beta_2$=0.129, $I=1/2$ and interval between the opposite parity levels $E(1/2^-)- E(1/2^+)$=55.2 KeV \cite{Spevak}.   Then  Eq.  (\ref{San}) gives:
       \begin{equation}\label{SThRa}
 S( ^{229}\textrm{Th})=  2 \, S( ^{225}\textrm{Ra}),
   \end{equation}     
Using $S( ^{225}\textrm{Ra})=300  \times 10^{-8} $ $e \eta$ fm$^3$ \cite{Spevak} we obtain  $S( ^{229}\textrm{Th})=600  \times 10^{-8} $ $e \eta$ fm$^3$. 

Within the meson exchange theory the $\pi$-meson exchange gives the dominating contribution to the T,P-violating nuclear forces \cite{SFK}. According to Ref. \cite{FDK} the neutron and proton constants in the T,P-odd potential  (\ref{eta}) may be presented as  $\eta_n \approx - \eta_p \approx 5 \times 10^6 (  -0.2 g {\bar g}_0 +  g {\bar g}_1 +  0.4 g {\bar g}_2$). In Refs. \cite{Auerbach,Spevak} we have not separated the proton and neutron contributions. Majority of the nucleons are neutrons, so it make sense to take $\eta=\eta_n$. However, the proton interaction constant has an opposite sign and may cancel a part of the neutron contribution, so we multiply the interaction constant by $((N-Z)/N)=0.36$ and use $\eta=0.36\eta_n$. This way we can obtain  a rough estimate: $S( ^{225}\textrm{Ra})=
(  - 2.2 g {\bar g}_0 +  11 g {\bar g}_1 + 4 g {\bar g}_2)\, e\, \textrm{fm}^3$,  $S(^{229}\textrm{Th})=
(  - 4.4 g {\bar g}_0 +  22 g {\bar g}_1 + 8 g {\bar g}_2)\, e\, \textrm{fm}^3$. 

A more accurate job has been done in Ref.  \cite{EngelRa}  where they presented the Schiff moment as 
$S( ^{225}\textrm{Ra})= (a_0 g {\bar g}_0 +  a_1 g {\bar g}_1 +  a_2 g {\bar g}_2) e $ fm$^3$. To estimate the error the authors  of Ref. \cite{EngelRa} have done the calculations using 4 different models of the strong interaction. They obtained the following 4 sets of the coefficients: $a_0= -1.5,\, -1.0,\, -4.7,\, -3.0$; 
   $a_1= 6.0,\, 7.0,\, 21.5,\, 16.9$;  $a_2= -4.0,\, -3.9,\, -11.0,\, -8.8$.    Taking the average values of the coefficients and using Eq. (\ref{SThRa}) we obtain:
  \begin{eqnarray}\label{SThRag}
 S( ^{225}\textrm{Ra})= (  - 2.6 g {\bar g}_0 +  12.9 g {\bar g}_1 -6.9 g {\bar g}_2)\, e\, \textrm{fm}^3,\\
 S( ^{229}\textrm{Th})= (  - 5.1 g {\bar g}_0 +  25.7 g {\bar g}_1 -13.9 g {\bar g}_2)\, e\, \textrm{fm}^3.
  \end{eqnarray}   
  We will use these expressions  as our final values for the Ra and Th  Schiff moments. We can express the results in terms of the more fundamental parameters such as the QCD $\theta$-term constant  ${\bar \theta}$ and the  quark chromo-EDMs ${\tilde d_u}$ and  ${\tilde d_d}$ using the relations $g {\bar g}_0=- 0.37{\bar \theta}$ \cite{Witten} and $g {\bar g}_0 = 0.8 \cdot10^{15}({\tilde d_u} +{\tilde d_d})$/cm,  $g {\bar g}_1 = 4 \cdot 10^{15}({\tilde d_u} - {\tilde d_d})$/cm  \cite{PR}:
  \begin{eqnarray}
  \label{SRatheta} S( ^{225}\textrm{Ra})= 1.0   \,{\bar \theta} \, e\, \textrm{fm}^3,\\
 \label{SThtheta}  S( ^{229}\textrm{Th})= 2.0  \, {\bar \theta} \, e\, \textrm{fm}^3,\\
 \label{SRaD}  S( ^{225}\textrm{Ra})= 10^4 ( 0.50 \,{\tilde d}_u -  0.54 \,{\tilde  d}_d )\, e\, \textrm{fm}^2,\\
  \label{SThD} S( ^{229}\textrm{Th})= 10^4 ( 1.0 \,{\tilde d}_u - 1.1\, {\tilde  d}_d )\, e\, \textrm{fm}^2.
 \end{eqnarray}   
  Note  that the contributions of ${\bar \theta}$ and ${\tilde d_{u,d}}$ should not be added to avoid double counting  since ${\tilde d_{u,d}}$ may be induced by ${\bar \theta}$.
   
 {\bf Molecular enhancement}.  Atomic EDM $d_a$ produced by the Schiff moment $S$ very rapidly increases with the nuclear charge $Z$ \cite{SFK,Khriplovich,KL,Spevak}:
 \begin{equation}\label{dZ}
 d_a \propto Z^2 (\frac{a_B}{2ZR})^{2- 2 \gamma} S,
 \end{equation}
 where $R$ is the nuclear radius, $a_B$ is the Bohr radius, $\gamma=\sqrt{1 - (Z \alpha)^2}$.  Th and Ra have close nuclear charges, $Z=88$ and $90$, and similar electronic structure up to last filled $7s^2$ subshell. Two extra $6d^2$ electrons in Th have high angular momenta, do not penetrate the nucleus and do not interact with the Schiff moment  directly (up to many-body corrections).  Therefore,  $d_a/S$ for Th is approximately equal to $d_a/S$ for Ra. Using calculations of Ra atom EDM from Refs. \cite{AtomicSchiff,AtomicSchiff2} we have
 \begin{equation}\label{dS}
 d_a(\textrm{Th}) \approx  -9 \cdot 10^{-17} \frac{S}{|e|\, \textrm{fm}^3} |e| \, \textrm{cm}= -2 \cdot 10^{-16} {\bar \theta}\, |e| \, \textrm{cm}.
 \end{equation}
$ d_a$(Th) as a function of other T,P and CP--violating interaction constants $\eta,\, {\bar g}, \, {\tilde d}$ can be found by the substitution of the Th Schiff moment from the equations  in the nuclear Schiff moment section above. This value of Th EDM is 3 orders of magnitude larger than Hg EDM and 4 orders of magnitude larger than Xe EDM. However, Th atom has non-zero electron angular momentum, $J=2$, and this reduces the signal coherence time and increases systematic errors. In principle, one may use Th$^{4+}$ ion which has closed shells or look for zero electron angular momentum Th ions in solid state materials. 

Note that the measurements of the effects produced by the  $^{229}$Th Schiff moment may be used to search for axions. Indeed, the  axion dark matter produces oscillating neutron EDM \cite{Graham} and oscillating  Schiff moment \cite{Stadnik}, the latter  is enhanced in $^{229}$Th by the same octupole mechaninism. Indeed, the axion dark matter field $a(t)=a_0 cos(m_a t)$ ($m_a$ is the axion mass)  generates oscillating nuclear forces which are similar to the T,P-odd nuclear forces producing the Schiff moments. To obtain the result for the oscillating Schiff moments and EDM it is sufficient to replace the constant ${\bar \theta}$ by $a(t)/f_a$, where $f_a$ is the axion decay constant \cite{Graham,Stadnik}.   Search for the effects produced by the oscillating axion-induced Schiff moments in solid state materials is in progress \cite{Casper}. A promising  direction here may be to use  $^{229}$ThO molecule placed in a matrix of Xe (or other) atoms. A proposal to use paramagnetic molecules in the matrix  of  rare-gas atoms for the electron  EDM search has been described in Ref. \cite{Kozlov}. 
 
    Promising objects for the Th Schiff moment measurement may be ThO molecule and ThOH$^+$ molecular ion. Both molecules have zero electron angular momentum in the ground state and very close opposite parity levels which enhance T,P-violating EDM.
    
    Use of polar diatomic  molecules for the measurement of the nuclear Schiff moment was suggested by  Sandars \cite{Sandars,Hinds} because electric field inside polarised molecule exceeds external electric field  $\epsilon $ by several  orders of magnitude and has the same direction.  The  molecular polarisation is $P \sim D\epsilon/(E_-  - E_+)$, where $D \sim e a_B$ is the intrinsic  electric dipole moment of the polar molecule. Therefore, to have a significant polarization degree $P$  the interval between the opposite parity molecular rotational levels $(E_-  - E_+)$ should be sufficiently  small.  Indeed, the rotational  interval in molecules is 3-5 orders of magnitude smaller than a typical interval between the opposite parity levels in atoms. 
    
    We may interpret the molecular enhancement in a different way \cite{SushkovFlambaum}: interaction between the Schiff moment and electrons mixes close opposite parity levels in the molecule, polarises the molecule along its angular momentum and creates T,P-violating EDM proportional to the large intrinsic electric dipole moment $D$ - see the discussion below Eq. (\ref{Scol}).  This enhanced EDM interacts with the external electric field  $\epsilon $.  The experiment has been performed with the TlF molecule \cite{TlFexperiment}. 
     In the paper  Ref. \cite{RaO} it was proposed to study molecule $^{225}$RaO where the effect may be 500 times larger than in TlF due to the enhanced Schiff moment and larger nuclear charge $Z$.  
    The best sensitivity to the electron EDM has  been obtained using molecules ThO \cite{TheEDM} and HfF$^+$ \cite{Cornell} in the excited metastable electronic state $^3\Delta_1$ which contains doublets of very close opposite parity levels. 
    Finally, in the recent paper \cite{MOH+} it was suggested that linear molecules MOH, molecular ions MOH$^+$ (M is a heavy atom, e.g. Ra in the molecule RaOH$^+$ ) and symmetric top molecules (such as MCH$_3$ or MOCH$_3$) may be better systems than molecules MO since such polyatomic molecules have a doublet of the close opposite parity energy levels in the bending mode and may be polarised by a weak electric field. The reduction of the  strength of the necessary electric field simplifies the experiment and dramatically reduces systematic effects.

    The T,P-violating effect in $^{229}$ThO is much larger than in TlF due to the enhanced Schiff moment and larger nuclear charge. An additional advantage may appear in ThOH$^+$ which is expected to have very close opposite parity states (similar to RaOH$^+$).  Another possibility may be  to use  the doublet in $^3\Delta_1$ metastable state of $^{229}$ThO (used  to improve the limit on electron EDM ) and the ground state  doublet  $^3\Delta_1$ in ThF$^+$.
    
    The interaction constant $W_S$ for the effective T,P-violating interaction in molecules 
 \begin{equation}\label{WS}   
   W_{T,P}=W_S\frac{S}{I} {\bf I \cdot n} 
 \end{equation}    
 (here ${\bf I}$ is the nuclear spin, ${\bf n}$ is  the  unit vector along the molecular axis) in ThO, ThOH$^+$  and RaOH$^+$  may be  estimated by the comparison with the RaO molecule. Calculation of $W_S$ for RaO has been done in Ref. \cite{RaOTitov}: $W_S(\textrm{RaO})=45192$ atomic units (here a.u.=$e/a_B^4$).
    In RaOH$^+$  ion the electron density on the Ra nucleus is slightly smaller than in RaO (since a part of the electron charge density moves to hydrogen), therefore, we assume  $W_S(\textrm{RaOH}^+) \approx 30000$ a.u. In ThO and in  ThOH$^+$ the electron density on the Th nucleus is expected to be slightly larger than that for Ra due to the higher Th charge and two extra electrons. Therefore, we assume  $W_S(\textrm{ThO}) \approx 50000$ a.u. and  $W_S(\textrm{ThOH}^+) \approx 30000$ a.u. Note that the electron wave function in the bending molecular mode of  RaOH$^+$ and ThOH$^+$ is the same as in their gound states, therefore, the parameters $W_S$ are practically not affected by these bending vibrations (where we have the doublet of the opposite parity levels).  The parameter $W_S$  in the $^3\Delta_1$ state   of the  $^{229}$ThO molecule   should have comparable value to their ground state values since  $^3\Delta_1$ and the ground $^1\Sigma_0$ state differ  by one electron orbital only.  The  $^3\Delta_1$ state in ThF$^+$  molecular ion is similar to the  $^3\Delta_1$ state in $^{229}$ThO molecule.
    
    The estimates presented above are based on comparison with the numerical calculations of the Schiff moment contribution in RaO. Estimates based on the $Z$ dependence extrapolation Eq. (\ref{dZ}) from TlF give 2 times larger results.
    
    Substitution of the Schiff moment (\ref{SThtheta}) to the energy shift $W_{T,P}=W_S\frac{S}{I} {\bf I \cdot n}$ gives for the fully polarised molecule the energy difference between the $I_z=I$ and $I_z=-I$ states in $^{229}$ThO:
 \begin{equation}\label{WTP} 
 2 W_S S=1. \cdot 10^7 {\bar \theta} \, \textrm{Hz}
 \end{equation}    
 A similar estimate is valid for  molecular ions ThF$^+$  and ThOH$^+$. The measured shift in the 1991 TlF experiment \cite{TlFexperiment} was $-0.13 \pm 0.22$ mHz.    The same sensitivity in the $^{229}$ThO, ThF$^+$   or ThOH$^+$  experiments would allow one to improve the current limit  $| {\bar \theta} | < 10^{-10}$ and also the limits on other fundamental parameters of the CP-violation theories such as the strength of T,P-violating potential $\eta$, the $\pi NN$ interaction constants ${\bar g}$ and the quark chromo-EDMs ${\tilde d}$.
       
 {\bf Comaparison with existing and proposed experiments}. We should  compare suggested experiments with   $^{229}$ThO, ThF$^+$  and  $^{229}$ThOH$^+$ molecules  with other existing and proposed experiments. The best limit on the nuclear Schiff moment has been obtained in the measurement of Hg EDM \cite{HgEDM}. However, there is a theoretical problem here: the most recent sophisticated calculation \cite{Engel} was not able to find out even the sign of the Hg Schiff moment, different interaction models give very different results. There are two reasons for this: firstly, the Schiff moment is determined by the charge distribution of the protons. However, it is directed along the nuclear spin which in  $^{199}$Hg is carried by the valence neutron, i.e. the Schiff moment in $^{199}$Hg is determined by the many-body effects which are harder to calculate. The second reason is in the formula for the Schiff moment defined by Eq. (\ref{S}). There are two terms of opposite sign in this formula which tend to cancel each other, the main term and the screening term (remind the reader that the screening term kills the nuclear EDM contribution to the atomic EDM).   If we do not know each term sufficiently accurately, the final sign and the magnitude of the Schiff moment are unknown.
  
    Recently the interest in EDM experiments has moved towards molecules where the effects are very strongly enhanced by the close rotational levels and very strong internal "effective electric field". For example, the limit on electron EDM in ThO and HgF$^+$ experiments have been improved by more than an order of magnitude in comparison with the atomic EDM experiments. The Tl nuclear Schiff moment has been measured in the TlF experiment \cite{TlFexperiment}.
   Similar to $^{199}$Hg,  calculations of the $^{203,205}$Tl Schiff moments suffer from the problem of the cancellation between two approximately equal terms  in Eq. (\ref{S}) and the problem of the nuclear core polarization contribution (since there is a strong  cancellation between the two terms in the valence proton contribution in Tl \cite{SFK,FKS1986,FDK}). 
        
    Actually, the interpretation of the TlF experiment \cite{TlFexperiment} was done in terms of the proton EDM. However, here we  probably have even a more serious problem (below  we will follow the discussion in Ref. \cite{FDK}). Firstly, calculations with different choices  of the strong interaction give  different signs and magnitudes of the Schiff moment $S_p$ induced by the proton EDM ( since we also have here the cancellation between the main and  the screening contributions). The authors of the molecular calculation \cite{Coveney}  selected the maximal value out of 4 numbers calculated by A. Brown (this maximal number  leads to the strongest limit on the proton EDM), and this value of $S_p$ was used in all other molecular calculations for TlF \cite{Parpia,Quiney,Petrov} (see also \cite{AtomicSchiff}). There was no such accuracy investigation for the proton EDM contribution to the Hg Schiff moment but naively we may expect that the accuracy is actually lower than in Tl since the valence nucleon in $^{199}$Hg is neutron. 
    
    The second problem is that in practically any model the contribution of the T,P-violating nuclear forces  to the Schiff moment  is 1-2 orders magnitude larger than the proton EDM contribution (the ratio is "model independent" since the $\pi NN$ interaction constant appears in both contributions and cancels out in the ratio).
     Therefore, to obtain the limit on the proton EDM we neglect much larger contribution of the P,T-odd nuclear forces. Thus, in the Particle Data tables the limit on the proton EDM is presented assuming that there are no other contributions to atomic and molecular EDM. However, if we wish to test CP-violation theories such limits on the proton EDM from Hg and Tl EDM  can hardly be used.    
 
   These theoretical problems do not exist for the collective Schiff moments in the nuclei with the octupole deformation. The second screening term  is very small in this case since it is proportional to a very small intrinsic dipole moment $D$ of the "frozen" nucleus. If the distributions of the neutrons and protons are the same, $D=0$. Thus, there is no cancellation and the intrinsic Schiff moment is proportional to the known electric  octupole moment (which may actually be measured using probabilities of the octupole transitions between the rotational levels).   Then the calculation is reduced to the expectation value of the T,P-odd interaction $<\Omega|W|\Omega>$, here  $|\Omega>$ is the ground state of the "frozen" deformed nucleus  \cite{Auerbach,Spevak}. Calculation of one expectation value looks  more reliable than the calculation of the  infinite sum 
  $\sum_n \frac{|n><n|W|0>}{E_0 -E_n}$ in  nuclei where there is no single dominating contribution.
   Thus, calculations of the collective Schiff moments  look more "clean" theoretically. More importantly, the collective Schiff moment is enhanced by 2-3 orders of magnitude.
    

{\bf Conclusion}.    We propose to search for the T,P and CP violating effects in the molecule $^{229}$ThO where the effects  are 2-3  orders of magnitude larger than in TlF due to the enahnced Schiff moment of the  $^{229}$Th nucleus and large  nuclear charge.  An additional advantage may be in $^{229}$ThOH$^+$ molecular ion,  in  $^3\Delta_1$ state  of the  $^{229}$ThO molecule and in $^{229}$ThF$^+$ molecular ion which have very close opposite parity energy levels and may be polarised by a weak electric field.  The $^3\Delta_1$ state   of the  $^{229}$ThO molecule has already been used to measure electron EDM. The enhanced effects in these molecules may also be used to search for axions. $^{229}$Th  lives 7917 years, may be produced in macroscopic quantities (as it is done for the medical applications) and is very well studied in numerous experiments.

    This work is supported by the Australian Research Council and Gutenberg Fellowship. I am grateful to M. Kozlov,  N. Hutzler, A. Palffy, Jun Ye, D. DeMiIle, H. Feldmeier, N. Minkov, A. Afanasiev, P. Ring and TACTICA collaboration for useful discussions.


\begin{thebibliography}{99}
\bibitem{PR} M. Pospelov, A. Ritz, Ann. Phys. (Amsterdam) {\bf318}, 119 (2005).
\bibitem{ERK} J. Engel, M.J. Ramsey-Musolf, U. van Kolck, Prog.Part.Nucl.Phys. {\bf 71}, 21 (2013).
\bibitem{Khriplovich} I.B. Khriplovich, Parity Nonconservation in
Atomic Phenomena (Gordon and Breach, Amsterdam, 1991).
\bibitem{KL}  I.B. Khriplovich, S.K. Lamoreaux, CP violation without strangeness. (Springer-Verlag, Berlin, 1997).
\bibitem{GF} Violation of fundamental symmetries in atoms and tests of unification
theories of elementary particles. J.S.M. Ginges, V.V. Flambaum.
 Phys. Rep.  {\bf 397}, 63 (2004).
\bibitem{Schiff} L.I. Schiff, Phys. Rev. {\bf 132}, 2194 (1963).
\bibitem{Sandars}P. G. H. Sandars, Phys.Rev.Lett. {\ bf 19}, 1396 (1967).
\bibitem{Hinds}  E. A. Hinds and P. G. H. Sandars, Phys.Rev.A {\bf 21}, 471 (1980).
\bibitem{SFK} On the possibility of investigation of P- and T-odd nuclear forces
in atomic and molecular experiments. O.P. Sushkov, V.V. Flambaum,
 I.B. Khriplovich.  Zh. Exp. Teor. Fiz. {\bf 87}, 1521 (1984) [ Sov. JETP {\bf 60}, 873  (1984)].
\bibitem{FKS1985} V.V. Flambaum, I.B. Khriplovich, O.P. Sushkov, Phys. Lett. B {\bf 162}, 213 (1985).
\bibitem{FKS1986} On the P- and T-nonconserving nuclear moments, V.V. Flambaum,
I.B. Khriplovich, O.P. Sushkov. Nucl. Phys. A {\bf 449}, 750, 1986.
\bibitem{FG} Nuclear Schiff moment and time invariance violation in atoms.
V.V. Flambaum and J.S.M. Ginges,  Phys. Rev. A {\bf 65}, 032113 (2002).
\bibitem{HH} W.C. Haxton, E.M. Henley, Phys. Rev. Lett. {\bf 51}, 1937 (1983).
\bibitem{F1994}  Spin hedgehog and collective magnetic quadrupole moments
induced by parity and time invariance violating interaction. V. V. Flambaum,
Phys. Lett. B {\bf 320}, 211 (1994).
\bibitem{Auerbach} Collective $T$- and $P$-odd electromagnetic moments in nuclei
with octupole deformations. N. Auerbach, V. V. Flambaum, and V. Spevak,
Phys. Rev. Lett. {\bf 76}, 4316 (1996).
\bibitem{Spevak} Enhanced T-odd P-odd electromagnetic moments in reflection
 assymmetric nuclei. V. Spevak, N. Auerbach, and V.V. Flambaum, Phys.
 Rev. C {\bf 56}, 1357 (1997).
\bibitem{HgEDM} Reduced Limit on the Permanent Electric Dipole Moment of $^{199}$Hg.
B. Graner, Y. Chen, E.G. Lindahl, and B.R. Heckel. Phys. Rev. Lett. 116, 161601 (2016); Phys. Rev. Lett. 119,
119901 (E) (2017). 
\bibitem{EngelRa}Time-reversal violating Schiff moment of 225Ra.
J. Engel, M. Bender, J. Dobaczewski, J. H. de Jesus, and P. Olbratowski
Phys. Rev. C {\bf 68}, 025501 (2003). 
\bibitem{FlambaumRa}  Enhancement of parity and time invariance violation
in the radium atom. V.V. Flambaum.  Phys. Rev. A60, R2611-2613 (1999).
\bibitem{DzubaRa} Calculation of parity and time invariance violation
in the radium atom. V.A. Dzuba, V.V. Flambaum, J.S.M. Ginges,
 Phys. Rev. A61, 062509-1 - 062509-10 (2000).
\bibitem{AtomicSchiff} Electric dipole moments of Hg, Xe,Rn,Ra,Pu
and TlF induced by the nuclear Schiff moment and limits on time-reversal
violating interactions.  V.A. Dzuba, V.V. Flambaum, J.S.M.
Ginges and M.G. Kozlov. Phys. Rev. A66,012111 (2002) .
\bibitem{AtomicSchiff2} Calculation of P,T-odd electric dipole moments for diamagnetic atoms
 $^{129}$Xe, $^{171}$Yb, $^{199}$Hg, $^{211}$Rn, and $^{225}$Ra
V.A. Dzuba, V.V. Flambaum, S.G. Porsev, Phys Rev A80,032120 (2009).  
\bibitem{Engel2000}  Nuclear octupole correlations and the enhancement of atomic time-reversal violation.
J. Engel, J.L. Friar, and A.C. Hayes, Phys. Rev. C {\bf 61}, 035502 (2000).
\bibitem{FZ} Enhancement of nuclear Schiff moment and time reversal violation
in atoms due to soft nuclear octupole vibrations. V.V. Flambaum,
 V.G. Zelevinsky. Phys. Rev. C {\bf 68}, 035502 (2003).
\bibitem{soft2} Nuclear Schiff moment in nuclei with soft octupole and quadrupole 
vibrations. N. Auerbach, V.F. Dmitriev, V.V. Flambaum, A. Lisetskiy,
R.A. Sen'kov, V.G. Zelevinsky,   Phys. Rev. C {\bf 74}, 025502 (2006).
\bibitem{SushkovFlambaum} Effects of parity nonconservation in diatomic molecules.  O.P. Sushkov,
 V.V.Flambaum.   Zh. Exp. Teor. Fiz.  {\bf 75}, 1208, (1978) [Sov.  JETP {\bf 48}, 608 (1978)].
\bibitem{RaEDM} First Measurement of the Atomic Electric Dipole Moment of $^{225}$Ra.
R.H. Parker, M.R. Dietrich, M.R. Kalita, N.D. Lemke, K.G. Bailey, M. Bishof, J.P. Greene, R.J. Holt, W. Korsch, Z.-T. Lu, P. Mueller, T.P. O'Connor, and J.T. Singh, Phys. Rev. Lett. 114, 233002 (2015).
\bibitem{Peik} Nuclear laser spectroscopy of the 3.5eV transition in Th$^{229}$, E. Peik, Chr.  Tamm, Europhys. Lett., 61, 181 (2003)
\bibitem{Thclock}  A Single-Ion Nuclear Clock for Metrology at the 19th Decimal Place
. C. J. Campbell, A. G. Radnaev, A. Kuzmich, V. A. Dzuba, V. V.
 Flambaum, and A. Derevianko, Phys. Rev. Lett.108, 120802 (2012). 
\bibitem{FlambaumTh} Enhanced effect of temporal variation of the fine structure constant
 and strong interaction in 229Th. V.V. Flambaum,  Phys. Rev. Lett. 97, 092502 (2006).
\bibitem{FlambaumTh1} Enhancing the effect of Lorentz invariance and Einstein's equivalence
 principle violation in  nuclei and atoms, V.V. Flambaum. Phys. Rev. Lett. 117, 072501 (2016).
 \bibitem{Tkalya} Proposal for a Nuclear Gamma-Ray Laser of Optical Range. E. V. Tkalya. Phys. Rev. Lett. 106, 162501 (2011).
 \bibitem{production} Energy Splitting of the Ground-State Doublet in the Nucleus 229Th. B. R. Beck, J. A. Becker, P. Beiersdorfer, G. V. Brown, K. J. Moody, J. B. Wilhelmy, F. S. Porter, C. A. Kilbourne, and R. L. Kelley.
Phys. Rev. Lett. 98, 142501 (2007).
\bibitem{Minkov} Reduced transition probabilities for gamma decay of the 7.8 eV isomer in $^{229}$Th. N. Minkov, A.  Palffy, Phys. Rev. Lett, 118, 212501(2017).
\bibitem{Engel} Fully self-consisten calculations of nuclear Schiff moments. S. Ban, J. Dobaczewski, J. Engel, and A. Shukla
Phys. Rev. C {\bf 82}, 015501 (2010).
\bibitem{FDK}  V. V. Flambaum, D. DeMille and M.G. Kozlov Phys. Rev. Lett.  {\bf 113}, 103003 (2014).
\bibitem{Witten} R.J. Crewther, P. di Vecchia, G. Veneziano, E. Witten. Phys. Lett. B {\bf 91}, 487 (1980).
\bibitem{Graham}  Axion  Dark-Matter Detection with Cold Molecules, P. W.  Graham  and  S.  Rajendran, Phys. Rev. D84, 055013 (2011).
\bibitem{Stadnik} Axion-induced effects in atoms, molecules and nuclei: Parity nonconservation, anapole moments, electric dipole moments, and spin-gravity and spin-axion momentum couplings. Y.V. Stadnik, V.V. Flambaum, Phys. Rev. D{\bf 89}, 043522 (2014).
\bibitem{Casper} Proposal for a Cosmic Axion Spin Precession Experiment (CASPEr).
Dmitry Budker, Peter W. Graham, Micah Ledbetter, Surjeet Rajendran, and Alexander O. Sushkov,
Phys. Rev. X 4, 021030 (2014) .
\bibitem{Kozlov} M.G. Kozlov, A. Derevianko. Phys. Rev. Lett. {\bf 97}, 063001 (2006).
\bibitem{TlFexperiment} D. Cho, K. Sangster, E.A. Hinds, Phys. Rev. A {\bf 44}, 2783 (1991).  
\bibitem{RaO}Electric dipole moments of actinide atoms and RaO molecule.
V.V. Flambaum,  Phys.Rev. A{\bf 77},024501 (2008).
\bibitem{RaOTitov} Calculation of P,T-odd interaction effect in 225RaO. A.D. Kudashov,
  A.N. Petrov, L.V. Skripnikov,  N.S. Mosyagin, A.V. Titov, and V.V. Flambaum,
 Phys. Rev. A{\bf 87}, 020102 (R)  (2013).  
\bibitem{MOH+} Precision measurement of time-reversal symmetry violation with laser-cooled polyatomic molecules. I. Kozyrev, N.R. Hutzler, Phys. Rev. Lett., {\bf 119}, 133002 (2017). 
\bibitem{TheEDM} J. Baron et al (ACME collaboration), Science {\bf 343}, 269 (2014). 
\bibitem{Cornell} William B. Cairncross, Daniel N. Gresh, Matt Grau, Kevin C. Cossel, Tanya S. Roussy, Yiqi Ni, Yan Zhou, Jun Ye, and Eric A. Cornell. Phys. Rev. Lett. {\bf 119}, 153001 (2017).
\bibitem{Coveney}  P. V. Coveney and P. G. H. Sandars, J. Phys. B16, 3727 (1983).
\bibitem{Parpia}  F. A. Parpia, J. Phys. B30, 3983 (1997).
\bibitem{Quiney}  H. M. Quiney, J. K. Laerdahl, K. Faegri, Jr., and T. Saue, Phys. Rev. A57, 920 (1998).
\bibitem{Petrov}  A. N. Petrov, N. S. Mosyagin, T. A. Isaev, A. V. Titov, V. F. Ezhov, E. Eliav, and U. Kaldor, Phys. Rev. Lett. 88, 073001 (2002).



\end{thebibliography}
\end{document}